\documentclass[12pt]{article} 
\usepackage{epsfig}

\begin{document}
\input amssym.def 
\input amssym
\hfuzz=5.0pt
%
%
%
%
\def\vec#1{\mathchoice{\mbox{\boldmath$\displaystyle\bf#1$}}
{\mbox{\boldmath$\textstyle\bf#1$}}
{\mbox{\boldmath$\scriptstyle\bf#1$}}
{\mbox{\boldmath$\scriptscriptstyle\bf#1$}}}
\def\mbf#1{{\mathchoice {\hbox{$\rm\textstyle #1$}}
{\hbox{$\rm\textstyle #1$}} {\hbox{$\rm\scriptstyle #1$}}
{\hbox{$\rm\scriptscriptstyle #1$}}}}
\def\operatorname#1{{\mathchoice{\rm #1}{\rm #1}{\rm #1}{\rm #1}}}
\chardef\ii="10
\def\widehat{\mathaccent"0362 }
\def\widetilde{\mathaccent"0365 }
\def\vphi{\varphi}
\def\vrho{\varrho}
\def\vtheta{\vartheta}
\def\ih{{\i\over\hbar}}
\def\hi{\frac{\hbar}{\i}}
\def\CD{{\cal D}}
\def\CE{{\cal E}}
\def\CH{{\cal H}}
\def\CL{{\cal L}}
\def\CP{{\cal P}}
\def\CV{{\cal V}}
\def\half{{1\over2}}
\def\bhalf{\hbox{$\half$}}
\def\viert{{1\over4}}
\def\halfvphi{\hbox{$\frac{\vphi}{2}$}}
\def\halfvphis{\hbox{$\frac{\vphi'}{2}$}}
\def\halfvphiss{\hbox{$\frac{\vphi''}{2}$}}
\def\bviert{\hbox{$\viert$}}
\def\hhbox#1#2{\hbox{$\frac{#1}{#2}$}}
\def\dfrac#1#2{\frac{\displaystyle #1}{\displaystyle #2}}
\def\intT{\ih\int_0^\infty\d\,T\,e^{\i ET/\hbar}}
\def\pathint#1{\int\limits_{#1(t')=#1'}^{#1(t'')=#1''}\CD #1(t)}
\def\hbarm{{\dfrac{\hbar^2}{2m}}}
\def\hbarmq{{\dfrac{\hbar^2}{2mq}}}
\def\mzwei{\dfrac{m}{2}}
\def\overh{\dfrac1\hbar}
\def\ihbar{\dfrac\i\hbar}
\def\intt{\int_{t'}^{t''}}
\def\tn{\tilde n}
\def\pmb#1{\setbox0=\hbox{#1}
    \kern-.025em\copy0\kern-\wd0
    \kern.05em\copy0\kern-\wd0
    \kern-.025em\raise.0433em\box0}
\def\pathintG#1#2{\int\limits_{#1(t')=#1'}^{#1(t'')=#1''}\CD_{#2}#1(t)}
\def\limN{\lim_{N\to\infty}}
\def\Norm{\bigg({m\over2\pi\i\epsilon\hbar}\bigg)}
\def\hbaram{{\hbar^2\over8m}}
\def\bbbr{{\rm I\!R}}                                
\def\bbbn{{\rm I\!N}}                                
\def\bbbz{{\mathchoice {\hbox{$\sf\textstyle Z\kern-0.4em Z$}}
{\hbox{$\sf\textstyle Z\kern-0.4em Z$}}
{\hbox{$\sf\scriptstyle Z\kern-0.3em Z$}}
{\hbox{$\sf\scriptscriptstyle Z\kern-0.2em Z$}}}}    
\def\bbbc{{\mathchoice {\setbox0=\hbox{\rm C}\hbox{\hbox
to0pt{\kern0.4\wd0\vrule height0.9\ht0\hss}\box0}}
{\setbox0=\hbox{$\textstyle\hbox{\rm C}$}\hbox{\hbox
to0pt{\kern0.4\wd0\vrule height0.9\ht0\hss}\box0}}
{\setbox0=\hbox{$\scriptstyle\hbox{\rm C}$}\hbox{\hbox
to0pt{\kern0.4\wd0\vrule height0.9\ht0\hss}\box0}}
{\setbox0=\hbox{$\scriptscriptstyle\hbox{\rm C}$}\hbox{\hbox
to0pt{\kern0.4\wd0\vrule height0.9\ht0\hss}\box0}}}}
\def\CP{{\cal P}}
\def\CQ{{\cal Q}}
\def\Ai{\operatorname{Ai}} 
\def\Cl{\operatorname{Cl}} 
\def\SU{\operatorname{SU}} 
\def\dt{\d t}
\def\d{\operatorname{d}}
\def\e{\operatorname{e}}
\def\i{\operatorname{i}}
\def\sn{\operatorname{sn}}
\def\cn{\operatorname{cn}}
\def\max{\operatorname{max}}
\def\DI{D_{\,\rm I}}
\def\DII{D_{\,\rm II}}
\def\3dDII{D_{\,3d-\rm II}}
\def\DIII{D_{\,\rm III}}
\def\DIV{D_{\,\rm IV}}
\def\KI{K_{\,\rm I}}
\def\KII{K_{\,\rm II}}
\def\KIII{K_{\,\rm III}}
\def\vphi{\varphi}
\def\tvphi{{\tilde\varphi}}
\def\tomega{{\tilde\omega}}
\def\ttau{{\tilde\tau}}
\def\hvphi{{\hat\varphi}}
\def\homega{{\hat\omega}}
\def\htau{{\hat\tau}}
\def\ps{\operatorname{ps}}
\def\Ps{\operatorname{Ps}}
\def\Si{\operatorname{Si}}
\def\energyldrei{\e^{-\i\hbar T(p^2+1)/2m}}
\def\ints{\int_0^{s''}}
\def\OO{\operatorname{O}}
\def\SO{\operatorname{SO}}
\def\fg{{\frak g}}
\def\fs{{\frak s}}
\def\fl{{\frak l}}
\def\gsl{\fg\fs\fl}
\def\operatorname#1{{\mathchoice{\rm #1}{\rm #1}{\rm #1}{\rm #1}}}
\def\bbbone{{\mathchoice {\rm 1\mskip-4mu l} {\rm 1\mskip-4mu l}
{\rm 1\mskip-4.5mu l} {\rm 1\mskip-5mu l}}}
\def\pathint#1{\int\limits_{#1(t')=#1'}^{#1(t'')=#1''}\CD #1(t)}
\def\pathints#1{\int\limits_{#1(0)=#1'}^{#1(s'')=#1''}\CD #1(s)}
 
\begin{titlepage}
\centerline{\normalsize DESY 06--140 \hfill ISSN 0418 - 9833}
\centerline{\hfill August 2006}

\vskip.3in
\message{TITLE:}
\begin{center}
{\bf\Large Path Integral Approach for Quantum Motion 
\\[3mm]
on Spaces of Non-constant Curvature
\\[3mm]
According to Koenigs}
\end{center}
\message{Path Integral Approach for for Quantum Motion 
on Spaces of Non-constant Curvature According to Koenigs}
\vskip.5in
\begin{center}
{\large Christian Grosche}
\vskip.1in
{\normalsize\em II.\,Institut f\"ur Theoretische Physik}
\vskip.05in
{\normalsize\em Universit\"at Hamburg, Luruper Chaussee 149}
\vskip.05in
{\normalsize\em 22761 Hamburg, Germany}
\end{center}
\vskip2in
\normalsize
\begin{quote}
Contribution to the ``XII. International Conference on Symmetry Methods 
in Physics'', July 3--8, Yerevan, Armenia.
\end{quote}

\vfill
\begin{center}
{\bf Abstract}
\end{center}
In this contribution I discuss a path integral approach for the quantum motion
on two-dimensional spaces according to Koenigs, for short ``Koenigs-Spaces''. 
Their construction is simple: One takes a Hamiltonian from 
two-dimensional flat space and divides it by a two-dimensional superintegrable 
potential. These superintegrable potentials are the isotropic singular 
oscillator, the Holt-potential, and the Coulomb potential. In all cases 
a non-trivial space of non-constant curvature is generated. We can study 
free motion and the motion with an additional superintegrable potential. 
For possible bound-state solutions we find in all three cases an equation of
eighth order in the energy $E$. The special cases of the Darboux spaces are
easily recovered by choosing the parameters accordingly.
\end{titlepage}
\normalsize 
 
\setcounter{page}{1}%
\setcounter{equation}{0}%
\section{Introduction}%
\message{Introduction}%
In this contribution I discuss the quantum motion on spaces of
non-constant curvature according to Koenigs \cite{KOENIGS}, 
which I will call for short ``Koenigs-spaces''. 
The construction of such a space is simple. One takes
a two-dimensional flat Hamiltonian, $\CH$, including some potential $V$, and
divides $\CH$ by a potential $f(x,y)$ ($x,y\in\bbbr^2$) such that this 
potential takes on the form of a metric:
\begin{equation}
\CH_{\rm Koenigs}=\frac{\CH}{f(x,y)}\enspace.
\end{equation}
Such a construction leads to a very rich structure, and attempts to classify
such systems are e.g. due to Kalnins et al.
\cite{KKPM,KKMb} and Daskaloyannis and Ypsilantis \cite{DASYPS}.
Simpler examples of such spaces are the Darboux spaces,
where one chooses the potential $f(x,y)$ in such a way that it depends 
only on one variable \cite{KalninsKMWinter}. 
Another choice consists whether one chooses for $f(x,y)$ some 
arbitrary potential (or some superintegrable
potential) and taking into account that the Poisson bracket structure of the 
observables makes up a reasonable simple algebra 
\cite{DASYPS,FMSUW,KalninsKMWinter}.

In previous publications we have analyzed the quantum motion on Darboux spaces
by means of the path integral \cite{GROas,GROPOe}. The path integral approach
\cite{FH,GRSh} serves as a powerful tool to calculate the 
propagator, respectively the Green function of the quantum motion in such a
space. In the present contribution I apply the path integral technique to
three kinds of Koenigs-spaces, where a specific two-dimensional
superintegrable potential \cite{GROPOa} is chosen. They are the
two-dimensional isotropic singular oscillator (Section II), 
the Holt-potential (section III) and the two-dimensional Coulomb-potential
(Section IV). Section V is devoted to a summary and a discussion of the
results achieved.


\setcounter{equation}{0}%
\section{Koenigs-Space with Isotropic Singular Oscillator}
\message{Koenigs-Space with Isotropic Singular Oscillator}%
We start with the first example, where we take for the metric term
\begin{eqnarray}
\d s^2&=& f_{I}(x,y)(\d x^2+ \d y^2)\enspace,\\
f_{I}(x,y)&=& \alpha
(x^2+y^2)+\frac{\beta}{x^2}+\frac{\gamma}{y^2}+\delta\enspace, 
\end{eqnarray}
and $\alpha,\beta,\gamma,\delta$ are constants.
The classical Hamiltonian and Lagrangian in $\bbbr^2$ with the isotropic
singular oscillator as the superintegrable potential have the form:
\begin{eqnarray}
\CL&=&\frac{m}{2}\Big((\dot x^2+\dot y^2)-\omega^2(x^2+y^2)\Big)
-\hbarm\bigg(\frac{k_x^2-\viert}{x^2}+\frac{k_y^2-\viert}{y^2}\bigg)\enspace,\\
\CH&=&\frac{p_x^2+p_y^2}{2m}+\frac{m}{2}\omega^2(x^2+y^2)
+\hbarm\bigg(\frac{k_x^2-\viert}{x^2}+\frac{k_y^2-\viert}{y^2}\bigg)\enspace.
\end{eqnarray}
Counting constants, there are seven independent constants:
$\alpha,\beta,\gamma,\delta$, and $\omega,k_x,k_y$. An eighth constant can
be added by adding a further constant $\tilde\delta$ into the potential
of the Hamiltonian. It will be omitted in the following.
The first Koenigs-space $\KI$ is constructed by considering
\begin{equation}
\CH_{\KI}=\frac{\CH}{f_{I}(x,y)}\enspace,
\end{equation}
hence for the Lagrangian (with potential)
\begin{equation}
\CL_{\KI}=\frac{m}{2}f_{I}(x,y)(\dot x^2+\dot y^2)
 -\frac{1}{f_{I}(x,y)}\left[\frac{m}{2}\omega^2(x^2+y^2)
 +\hbarm\left(\frac{k_x^2-\viert}{x^2}+\frac{k_y^2-\viert}{y^2}\right)\right]
\enspace.
\end{equation}
Setting the potential in square-brackets equal to zero yields the Lagrangian
for the free motion in $\KI$. With this information we can set up the path
integral in $\KI$ including a potential. Because the space is two-dimensional,
and the metric is diagonal, the additional quantum potential $\propto\hbar^2$
vanishes. The canonical momentum operators are constructed by
\begin{equation}
p_{x_i}=\hi\bigg(\frac{\partial}{\partial_{x_i}}
+\frac{\Gamma_i}{2}\bigg)\enspace,\qquad
\Gamma_i=\frac{\partial}{\partial_{x_i}}\ln\sqrt{g}\enspace,
\end{equation}
with $x_1=x,x_2=y$ and $g=\det(g_{ab})$, $(g_{ab})$ the metric tensor.
For the path integral in the product lattice definition \cite{GRSh} we obtain
\begin{eqnarray}
&&
K^{(\KI)}(x'',x',y'',y';T)=\pathint{x}\pathint{y}f_{I}(x,y)
\nonumber\\  && 
=\exp\Bigg(\ih\intt\Bigg\{\frac{m}{2}f_{I}(x,y)(\dot x^2+\dot y^2)
\nonumber\\  &&\qquad\qquad\qquad\qquad
 -\frac{1}{f_{I}(x,y)}\Bigg[\frac{m}{2}\omega^2(x^2+y^2)
 +\hbarm\Bigg(\frac{k_x^2-\viert}{x^2}+\frac{k_y^2-\viert}{y^2}\Bigg)\Bigg]
 \Bigg\}\dt\Bigg)\qquad
\nonumber\\  &&
G^{(\KI)}(x'',x',y'',y';E)=\ih\int_0^\infty \d s'' K^{(\KI)}(x'',x',y'',y';s'')
                   \e^{\i\delta Es''/\hbar}\enspace,
\label{Green-integration}
\end{eqnarray}
with the time-transformed path integral $K^{(\KI)}(s'')$ given by
($\widetilde\omega^2=\omega^2-2\alpha E/m$)
\begin{eqnarray}
&&
K^{(\KI)}(x'',x',y'',y';s'')  
\nonumber\\  &&  
=\pathints{x}\pathints{y}
\exp\Bigg\{\ih\ints\Bigg[\frac{m}{2}
   \Big((\dot x^2+\dot y^2)-\widetilde\omega^2(x^2+y^2)\Big)
\nonumber\\  &&\qquad\qquad\qquad\qquad
   -\hbarm\Bigg(\frac{k_x^2-2m\beta E/\hbar^2-\viert}{x^2}
               +\frac{k_y^2-2m\gamma E/\hbar^2-\viert}{y^2}\Bigg)\Bigg]
   \d s''\Bigg\}\enspace.\qquad\qquad
\label{Ks-KI}
\end{eqnarray}
The path integrals in the variables $x$ and $y$ are both path integrals
for the radial harmonic oscillator, however with energy-dependent
coefficients. By switching to two-dimensional polar coordinates
$x=r\cos\vphi$, $y=r\sin\vphi$, the path integral in $x,y$ gives one
in $r,\vphi$. Furthermore, we get $x^2+y^2=r^2$,
$1/x^2=1/r^2\cos^2\vphi$, and  $1/y^2=1/r^2\sin^2\vphi$. Let us 
abbreviate $\tilde k_x^2=k_x^2-2m\beta E/\hbar^2$, 
$\tilde k_y^2=k_y^2-2m\gamma E/\hbar^2$. In the variable $\vphi$ we obtain a
path integral for the P\"oschl--Potential, and in the variable $r$ a 
radial path integral. The successive path integrations therefore
yield
\begin{eqnarray}
&&K^{(\KI)}(r'',r',\vphi'',\vphi';s'')
=\sum_{n_\vphi}\Phi_{n_\vphi}^{(\tilde k_y,\tilde k_x)}(\vphi'')
               \Phi_{n_\vphi}^{(\tilde k_y,\tilde k_x)}(\vphi')
\nonumber\\  &&\qquad\qquad\times
\frac{m\widetilde\omega\sqrt{r'r''}}{\i\hbar\sin\widetilde\omega s''}
\exp\Bigg[-\frac{m\widetilde\omega}{2\i\hbar}({r'}^2+{r''}^2)
       \cot\widetilde\omega s''\Bigg]
I_\lambda\Bigg(\frac{m\widetilde\omega r'r''}
                    {\i\hbar\sin\widetilde\omega s''}\Bigg)\enspace.
\end{eqnarray}
Here $\lambda=2n_\vphi+\tilde k_x+\tilde k_y+1$, and the 
$\Phi_{n_\vphi}^{(\tilde k_y,\tilde k_x)}(\vphi)$ are the wave-functions for
the P\"oschl-Teller potential \cite{BJb,GRSh}. $I_\lambda(z)$ is the
modified Bessel function \cite{GRA}. Performing the $s''$-integration for
obtaining  the Green function $G(E)$ yields \cite{GRA,GRSh}:
\begin{eqnarray}
&&G^{(\KI)}(r'',r',\vphi'',\vphi';E)
=\sum_{n_\vphi}\Phi_{n_\vphi}^{(\tilde k_y,\tilde k_x)}(\vphi'')
               \Phi_{n_\vphi}^{(\tilde k_y,\tilde k_x)}(\vphi')
\nonumber\\  &&\qquad\qquad\times
\frac{\Gamma\big[\half(1+\lambda-\delta E/\hbar\widetilde\omega)\big]}
     {\hbar\widetilde\omega\sqrt{r'r''}\,\Gamma(1+\lambda)}
W_{\delta E/2\widetilde\omega,\lambda/2}
  \bigg(\frac{m\widetilde\omega}{\hbar}r_>^2\bigg)
M_{\delta E/2\widetilde\omega,\lambda/2}
  \bigg(\frac{m\widetilde\omega}{\hbar}r_<^2\bigg)\enspace.\qquad
\label{Green-KI}
\end{eqnarray}
$M_{\mu,\nu}(z)$ and $W_{\mu,\nu}(z)$ are Whittaker-functions \cite{GRA}, and 
$r_<,r_>$ is the smaller/larger of $r',r''$.
The poles of the $\Gamma$-function give the energy-levels of the bound states:
\begin{equation}
\bhalf(1+\lambda-\delta E/\hbar\widetilde\omega)=-n_r\enspace,
\end{equation}
which is equivalent to ($N=n_r+n_\vphi+1=1,2,\dots$):
\begin{eqnarray}
\delta E  
&=&\hbar\sqrt{\omega^2-\frac{2\alpha}{m}E}
    \left(2N+\sqrt{k_x^2-\frac{2m\beta}{\hbar^2}E} \,
        +\sqrt{k_y^2-\frac{2m\gamma}{\hbar^2}E}\,\right)\enspace.
\end{eqnarray}
In general, this quantization condition is an equation of eighth order in $E$.
If we know the bound state energy $E_N$, we can determine the
wavefunctions according to
\begin{equation}
\Psi_N^{(\KI)}(r,\phi)=N_N\Phi_{n_\vphi}^{(\tilde k_y,\tilde k_x)}(\vphi)
\Phi_{n_r}^{(RHO,\lambda)}(r)\enspace,
\end{equation}
with the normalization constant $N_N$ determined by evaluating the residuum 
in the Green function (\ref{Green-KI}), and the $\Phi_N^{(RHO,\lambda)}(r)$
are the wave-functions of the radial harmonic oscillator \cite{GRSh}.
We can recover the flat space limit with $\alpha=\beta=\gamma=0$ with the
correct spectrum $E_N=\hbar\omega(N+k_x+k_y)/\delta$.

Note that we also can obtain the quantization condition by explicitly
inserting the wave-functions in $x$ and $y$ in (\ref{Ks-KI}) and performing
the $s''$-integration in (\ref{Green-integration}). 
We do not discuss the continuous spectrum.


\setcounter{equation}{0}%
\section{Koenigs-Space with Holt-Potential}
\message{Koenigs-Space with Holt-Potential}
Next we consider for the metric term
\begin{eqnarray}
\d s^2&=& f_{II}(x,y)(\d x^2+ \d y^2)\enspace,\\
f_{II}(x,y)&=&\alpha(x^2+4y^2)+\frac{\beta}{x^2}+\gamma y+\delta\enspace
\end{eqnarray}
and $\alpha,\beta,\gamma,\delta$ are constants.
The classical Hamiltonian and Lagrangian in $\bbbr^2$ with the 
Holt-potential as the superintegrable potential have the form:
\begin{eqnarray}
\CL&=&\frac{m}{2}\Big((\dot x^2+\dot y^2)-\omega^2(x^2+4y^2)\Big)
 -k_y y-\hbarm\frac{k_x^2-\viert}{x^2}\enspace,\\
\CH&=&\frac{p_x^2+p_y^2}{2m}+\frac{m}{2}\omega^2(x^2+4y^2)
+k_y y+\hbarm\frac{k_x^2-\viert}{x^2}\enspace.
\end{eqnarray}
Counting constants, there are seven independent constants:
$\alpha,\beta,\gamma,\delta$, and $\omega,k_x,k_y$. An eighth constant can
be added by adding a further constant $\tilde\delta$ into the potential
of the Hamiltonian, which is omitted. The second Koenigs-space $\KII$ with
potential is now constructed by considering 
\begin{equation}
\CH_{\KI}^{(V)}=\frac{\CH}{f_{II}(x,y)}\enspace.
\end{equation}
From the discussion in the Section II it is obvious how to construct the
path integral on $\KII$. We proceed straightforward to the time-transformed
path integral $K^{(\KII)}(s'')$ which has the form
\begin{eqnarray}
&&\!\!\!\!\!\!\!\!
K^{(\KII)}(x'',x',y'',y';s'')=\pathints{x}\pathints{y}
\nonumber\\  &&\!\!\!\!\!\!\!\!\qquad\times
\exp\Bigg\{\ih\ints\Bigg[\frac{m}{2}
   \Big((\dot x^2+\dot y^2)-\widetilde\omega^2(x^2+4y^2)\Big)
   -\hbarm\frac{\tilde k_x^2-\viert}{x^2}-(k_y-\gamma E)y
  \Bigg]\d s''\Bigg\}\,.\qquad
\end{eqnarray}
Again, $\widetilde\omega^2=\omega^2-2\alpha E/m$, 
$\tilde k_x^2=k_x^2-2m\beta E/\hbar^2$.
We have in the variable $x$ a singular oscillator, and in the variable $y$
a shifted oscillator with shift $y\to y-(k_y-\gamma E)/
(4m\widetilde\omega^2)\equiv y-y_E$. 
However, in comparison to Section II, we cannot 
separate variables in an analogous way as for $\KI$, because the only
separating coordinate systems for the Holt-potential are the Cartesian and
the parabolic systems, and only in Cartesian coordinates a closed solution is
possible. Therefore we must evaluate this path integral by another method.
The first possibility consists of writing down the Green functions for the
radial singular oscillator $G^{(RHO,\tilde k_x)}(E)$ and for the shifted
harmonic oscillator $G^{(HO,y_E)}(E)$, respectively. These solutions can be
found in e.g.~\cite{GRSh}. The final result for the Green function 
$G^{(\KII)}(E)$ then has the form
\begin{equation}
G^{(\KII)}(E)=\frac{\hbar}{2\pi\i}\int\d\CE 
  G_x^{(RHO,\tilde k_x)}(E;x'',x';\CE) G_y^{(HO,y_E)}
  \bigg(E;y'',y';-\CE-\delta
   +\frac{(k_y-\gamma E)^2}{8m\widetilde\omega^2}\bigg)\enspace.
\label{Green-KII}
\end{equation}
However, this is a very complicated expression, mainly due to the fact that
both the Green functions $G^{(RHO)}(E)$ and $G^{(HO,{\rm shift})}(E)$ consist
of products of Whittaker functions and parabolic cylinder functions,
respectively. A better way to analyze the spectral properties is to re-express
each kernel in its bound-state wave-functions expansion. Therefore
\begin{eqnarray}
&&\!\!\!\!\!\!\!\!
K^{(\KII)}(x'',x',y'',y';s'')
=\sum_{n_x}\Psi_{n_x}^{(RHO,\tilde k_x)}(x'')\Psi_{n_x}^{(RHO,\tilde k_x)*}(x')
\sum_{n_y}\Psi_{n_y}^{(HO,y_E)}(y'')\Psi_{n_y}^{(HO,y_E)*}(y')
\nonumber\\  &&\qquad\qquad\qquad\qquad\qquad\qquad\times
\e^{-\i s''(k_y-\gamma E)^2/(8m\hbar\widetilde\omega^2)}
\e^{-\i s''\widetilde\omega(n_x+\tilde k_x+2n_y+3/2)}\enspace.
\end{eqnarray}
Here, the $\Psi_{n_y}^{(HO,y_E)}(y)$ denote the wave-functions of a shifted
harmonic oscillator with shift $y_E$.
Performing the $s''$-integration similarly as in (\ref{Green-integration})
we get the quantization condition ($N=n_x+2n_y+3/2$)
\begin{equation}
8m\delta E\bigg(\omega^2-\frac{2\alpha}{m}E\bigg)-(k_y-\gamma E)^2
=\hbar \bigg(\omega^2-\frac{2\alpha}{m}E\bigg)^{3/2}
\left(2N+\sqrt{k_x^2-\frac{2m\beta}{\hbar^2}E}\,\right)\enspace. 
\label{Energy-KII}
\end{equation}
In general, this is an equation of eighth order in $E$.
The solution in terms of the wave-functions then has the form
\begin{equation}
\Psi_N^{(\KII)}(x,y)=
N_N\Psi_{n_x}^{(RHO,\tilde k_x)}(x)\Psi_{n_y}^{(HO,y_E)}(y)\enspace,
\end{equation}
and the normalization constant $N_N$ is determined by the residuum
of (\ref{Green-KII}) at the energy $E_N$ from (\ref{Energy-KII}).
The correct flat space limit with $\alpha=\beta=\gamma=0$ is easily recovered
with spectrum $E_N=\hbar\omega (N+k_x)/\delta+k_y^2/8m\delta\omega^2$.
We do not discuss the continuous spectrum. 

\setcounter{equation}{0}%
\section{Koenigs-Space with Coulomb-Potential}
\message{Koenigs-Space with Coulomb-Potential}
For the last example we consider a metric which corresponds to the
two-dimensional Coulomb potential ($r^2=x^2+y^2$)
\begin{eqnarray}
\d s^2&=& f_{III}(x,y)(\d x^2+ \d y^2)\enspace,\\
f_{III}(x,y)&=&-\frac{\alpha_1}{r}
   +\frac{1}{4r^2}\Bigg(\frac{\beta}{\cos^2\halfvphi}
   +\frac{\gamma}{\sin^2\halfvphi}\Bigg)+\delta
\end{eqnarray}
and $\alpha_1,\beta,\gamma,\delta$ are constants.
The classical Hamiltonian and Lagrangian in $\bbbr^2$ with the 
Coulomb potential as the superintegrable potential have the form:
\begin{eqnarray}
\CL&=&\frac{m}{2}(\dot x^2+\dot y^2)
 +\frac{\alpha_2}{r}
   -\frac{\hbar^2}{8mr^2}\Bigg(\frac{k_1^2-\viert}{\cos^2\halfvphi}
   +\frac{k_2^2-\viert}{\sin^2\halfvphi}\Bigg)\enspace,\\
\CH&=&\frac{p_x^2+p_y^2}{2m}-\frac{\alpha_2}{r}
   +\frac{\hbar^2}{8mr^2}\Bigg(\frac{k_1^2-\viert}{\cos^2\halfvphi}
   +\frac{k_2^2-\viert}{\sin^2\halfvphi}\Bigg)\enspace.
\end{eqnarray}
Counting constants, there are seven independent constants:
$\alpha_1,\beta,\gamma,\delta$, and $\alpha_2,k_1,k_2$. 
An eight constants can be added by adding a further constant 
$\tilde\delta$ into the potential of the Hamiltonian, which is again omitted.
The third Koenigs-space $\KIII$ is constructed by considering
\begin{equation}
\CH_{\KI}^{(V)}=\frac{\CH}{f_{III}(x,y)}\enspace.
\end{equation}
We proceed to the time-transformed path integral
$K^{(\KIII)}(s'')$ which has the form
\begin{eqnarray}
&&
K^{(\KIII)}(r'',r',\vphi'',\vphi';s'')=\pathints{r}\pathints{\vphi}r
\nonumber\\  &&\qquad\times
\exp\Bigg\{\ih\ints\Bigg[\frac{m}{2}(\dot r^2+r^2\dot\vphi^2)
-\frac{\tilde\alpha}{r}
  -\frac{\hbar^2}{8mr^2}\Bigg(\frac{\tilde k_1^2-\viert}{\cos^2\halfvphi}
               +\frac{\tilde k_2^2-\viert}{\sin^2\halfvphi}-1\Bigg)\Bigg]
   \d s''\Bigg\}\,.\qquad
\end{eqnarray}
Here, $\tilde k_1^2=k_1^2-2m\beta E/\hbar^2$, 
$\tilde k_2^2=k_2^2-2m\gamma E/\hbar^2$,
$\tilde\alpha=\alpha_2-\alpha_1 E$.
As in Section II, it is best to switch to two-dimensional polar coordinates,
which is straightforward. We obtain for the Green function in polar coordinates
\begin{eqnarray}
&&
G^{(\KIII)}(r'',r',\vphi'',\vphi';E)
=\sum_{n_\vphi}\Phi_{n_\vphi}^{(\tilde k_2,\tilde k_1)}(\halfvphiss)
\Phi_{n_\vphi}^{(\tilde k_2,\tilde k_1)*}(\halfvphis)
\nonumber\\  &&\qquad\times
\frac{1}{\hbar}\sqrt{-\frac{m}{2\delta E}}\,
\frac{\Gamma\big(\half+\lambda-\kappa\big)}{\Gamma(2\lambda+1)}
W_{\kappa,\lambda}\left(\sqrt{-8m \delta E}\,\frac{r_>}{\hbar}\right)
M_{\kappa,\lambda}\left(\sqrt{-8m \delta E}\,\frac{r_>}{\hbar}\right)
\label{Green-KIII}
\end{eqnarray}
$\Big(\kappa=(\tilde\alpha/\hbar)\sqrt{-m/2\delta E}$,
$\lambda=n_\vphi+\tilde k_1/2+\tilde k_2/2+\half\Big)$.
The poles of the $\Gamma$-function gives the quantization condition
$1/2+\lambda-\kappa=-n_r$, or more explicitly
\begin{equation}
1+n_\vphi+n_r+\half\sqrt{k_1^2-\frac{2m\beta}{\hbar^2}E}\,
             +\half\sqrt{k_2^2-\frac{2m\gamma}{\hbar^2}E}
=\frac{\alpha_2-\alpha_1 E}{\hbar}\sqrt{-\frac{m}{2\delta E}}\enspace.
\label{Energy-KIII}
\end{equation}
This is again an equation of eighth order in $E$.
Actually, this quantization condition has the same structure as the 
quantization condition for the third potential on Darboux Space $\DII$,
c.f.~our recent publication \cite{GROPOe}. We consider the special case
$k_1=k_2=0$. This gives ($N=1+n_\vphi+n_r$):
\begin{equation}
N=\frac{\alpha_2-\alpha_1 E}{\hbar}\sqrt{-\frac{m}{2\delta E}}
 -\frac{\sqrt{-E}}{2\hbar}\Big(\sqrt{2m\beta}+\sqrt{2m\gamma}\,\Big)\enspace.
\end{equation}
This is a quadratic equation in the energy $E$ with solution
\begin{eqnarray}
&&E_{\pm}=-\frac{B}{2A}\pm\frac{1}{2A}\sqrt{B^2-4AC}\enspace,
\\   
&&\left.
\begin{array}{l}
A=m\alpha_1(a_1-2)+2m\delta\left(\sqrt{\beta}+\sqrt{\gamma}\,\right)^2\enspace,
\\
B=2\delta\hbar^2 N^2+2\alpha_2(m-\alpha_1)\enspace,\qquad
C=m\alpha_2^2\enspace.
\end{array}\qquad\right\}\qquad\qquad
\end{eqnarray}
We consider the limit $N\to\infty$. In this case, we take the $+$-sign of
the square-root expression only, and obtain
\begin{equation}
E_N\simeq -\frac{m\alpha_2^2}{2\delta\hbar^2 N^2}\enspace,\qquad 
(N\to\infty)\enspace,
\end{equation}
showing a Coulomb-behavior of the energy-levels.
For the bound-states wave-function we get in the general case
($a=\hbar^2/m\tilde\alpha$):
\begin{eqnarray}
&&
\Psi_N^{(\KIII)}(r,\vphi)=
\frac{N_N}{n_r+\lambda+\half}\sqrt{\frac{n_r!}{a\Gamma(n_r+2\lambda+1)}}\,
\Phi_{n_\vphi}^{(\tilde k_2,\tilde k_1)}(\halfvphi)
\nonumber\\  &&\qquad\times
\Bigg(\frac{2r}{a(n_r+\lambda+\half)}\Bigg)^\lambda
\exp\Bigg[-\frac{r}{a(n_r+\lambda+\half)}\Bigg]
L_{n_r}^{(2\lambda)}\Bigg(\frac{2r}{a(n_r+\lambda+\half)}\Bigg)
\end{eqnarray}
(the $L_n^{(\lambda)}(z)$ are Laguerre polynomials \cite{GRA}).
The wave-functions in $r$ are the well-known Coulomb wave-functions.
Note that $\lambda=\lambda(E_N)$.
The normalization constant $N_N$ is determined by taking the residuum
in the Green function (\ref{Green-KIII}) for the corresponding 
energy $E_N$ from (\ref{Energy-KIII}). 

We get another special case if we set the potential in $\KIII$ to zero,
i.e., $k_{1,2}=\half,\alpha_2=0$. This yields together with the simplification
$\beta=\gamma$
\begin{equation}
N+\sqrt{\viert-\frac{2m\beta E}{\hbar^2}}=-\frac{\alpha_1 E}{\hbar}
\sqrt{-\frac{m}{2\delta E}}\enspace.
\end{equation}
This is a quadratic equation in the energy $E$ with solution
\begin{eqnarray}
&&E_{\pm}=-\frac{B}{2A}\pm\frac{B}{2A}\sqrt{1-\frac{4AC}{B^2}}\enspace,
\\   
&&\left.\begin{array}{l}
A=\frac{m^2}{\hbar^4}\bigg(\frac{\alpha_1^2}{2\delta}-4\beta N\bigg)^2
\enspace,\qquad
C=(N^2+N)^2-4N^2\enspace,
\\[4mm]
B=\frac{2m}{\hbar^2}\Bigg[(N^2+N)
     \bigg(\frac{\alpha_1^2}{2\delta}-4\beta N\bigg)+8\beta\Bigg]\enspace.
\end{array}\qquad\right\}
\end{eqnarray}
We see that even for zero potential, bound states are possible.
For $N\to\infty$, the leading term behaves according to $-B/2A\to
\hbar^2N/2m\beta$, showing a oscillator-like behavior.
We do not discuss the continuous
spectrum. This concludes the discussion. 
 
 
\setcounter{equation}{0}%
\section{Summary and Discussion}
\message{Summary and Discussion}
In this contribution I have discussed a path integral approach for spaces of
non-constant curvature according to Koenigs, which I have for short called
``Koenigs-spaces'' $\KI$, $\KII$, and $\KIII$, respectively.
I have found a very rich structure of the spectral properties of the quantum
motion on Koenigs-spaces. In the general case with potential, 
in all three spaces the
quantization condition is determined by an equation of eighth order in the
energy $E$. Such an equation cannot be solved explicitly, however special
cases can be studied. Indeed in the space $\KIII$ we have found for such a
special case a Coulomb-like spectrum for large quantum numbers. This is very
satisfying, because the flat space $\bbbr^2$ is contained as a special case of 
$\KIII$. Our systems are also superintegrable, because they admit separation
of variables in more that one coordinate system.

Let us note a further feature of these spaces. It is obvious that our solutions
remain on a formal level. Neither have we specified an embedding
space, nor have we specified boundary conditions on our spaces.
Let us consider the space $\KII$: We set $\alpha=\beta=\delta =0$ and
$\gamma=1$. In this case we obtain a metric which corresponds to the Darboux
space $\DI$ (modulo change of variables), as discussed in
\cite{KalninsKMWinter}. In $\DI$ boundary 
conditions and the signature of the ambient space is very important, because
choosing a positive or a negative signature of the ambient space changes
the boundary conditions, and hence the quantization conditions 
\cite{GROPOe}.

Furthermore, we can recover the Darboux space $\DII$
\cite{GROas,GROPOe,KalninsKMWinter} by setting in our examples in the
potential function $f$ all constant to zero except those corresponding 
to the $1/x^2$-singularity. However, we did not discuss these cases in detail.

In our approach we have chosen examples of superintegrable potentials
in two-di\-men\-sional space, i.e. the isotropic singular oscillator, the Holt
potential and the Coulomb potential, respectively. 
Other well-known potentials can also be included, for instance the 
Morse-potential or the (modified) P\"oschl--Teller potential. Actually, the
incorporation of the Morse-potential leads to the Darboux space $\DIII$, and
the incorporation of the P\"oschl--Teller potential to the Darboux space 
$\DIV$ \cite{KalninsKMWinter}. The quantum motion without potential have been
discussed extensively in \cite{GROas}, and with potentials will be discussed
in \cite{GROPOf}. In these cases, also complicated quantization conditions
are found.

In the present contribution I have omitted the discussion of the continuous
spectrum. This is on the one hand side due to lack of space, and on the other
the specific ambient space has to be taken into account. For instance, in the
Darboux space $\DII$ we know that the continuous spectrum
has the form of $E_p\propto(\hbar^2/2m)p^2+\hbox{constant}$. The
wave-functions are proportional to K-Bessel functions \cite{GROas}.
However, in Darboux space $\DI$ there is no such constant, and the
wave-functions have a different form. Furthermore, $\DII$ contains as special
cases  the two-dimensional Euclidean plane and the Hyperbolic plane,
respectively. In $\KII$ we can find these spaces for a special
choice of parameters and the continuous wave-functions are proportional
to Whittaker-functions (which reduce to K-Bessel functions and parabolic
cylinder functions for specific parameters, respectively). 
Such a more detailed study will be presented elsewhere.

\subsection*{\bf Acknowledgments}
This work was supported by the Heisenberg--Landau program. 

The author is grateful to Ernie Kalnins for fruitful and pleasant
discussions on superintegrability and separating coordinate systems. 
I also would like to thank the organizers, in particular G.Pogosyan,  
of the ``XII. International Conference on Symmetry Methods in Physics'', 
July 3--8, Yerevan, Armenia, for the warm hospitality during my stay in
Yerevan. 


\bigskip\bigskip 
\vbox{\centerline{\ }
\centerline{\quad\epsfig{file=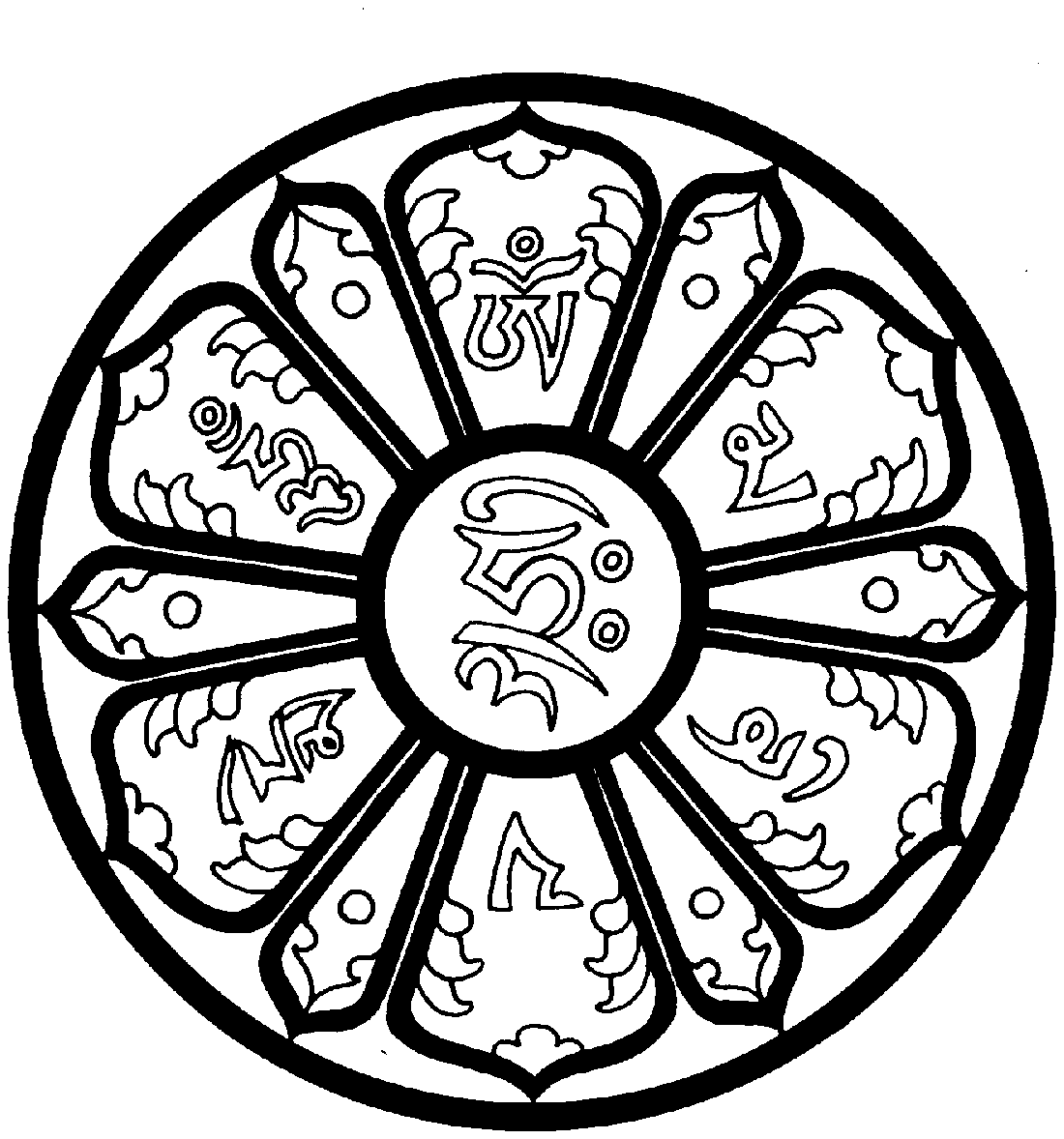,width=4cm}}}

\end{document}